\documentstyle[prb,aps,ams,epsfig]{revtex}
\newcommand \n{\noindent}
\newcommand \beq{\begin{eqnarray}}
\newcommand \eeq{\end{eqnarray}}
\newcommand \be{\begin{eqnarray}}
\newcommand \ee{\end{eqnarray}}

\newcommand{\set}[2]{\newcommand{#1}{#2}}
\set{\pa}{\partial \over \partial\, }
\set{\leftvector}{\stackrel{\leftarrow}{\partial }}
\set{\rightvector}{\stackrel{\rightarrow}{\partial }}
\begin{document}
\twocolumn[\hsize\textwidth\columnwidth\hsize
           \csname @twocolumnfalse\endcsname
\title{
Conductivity in quasi two-dimensional systems}
\author{K. Morawetz}
\address{Max-Planck-Institute for the Physics of Complex Systems,
Noethnitzer Str. 38, 01187 Dresden, Germany
}
\maketitle
\begin{abstract}
The conductivity in quasi two-dimensional systems is calculated using the quantum kinetic equation. Linearizing the Lenard-Balescu collision integral with the extension to include external field dependences allows one to calculate the conductivity with diagrams beyond the GW approximation including maximally crossed lines. Consequently the weak localization correction as an interference effect appears here from the field dependence of the collision integral (the latter dependence sometimes called intra-collisional field effect). It is shown that this weak localization correction has the same origin as the Debye-Onsager relaxation effect in plasma physics. The approximation is applied to a system of quasi two-dimensional electrons in hetero-junctions which interact with charged and neutral impurities and the low temperature correction to the conductivity is calculated analytically. It turns out that the dynamical screening due to charged impurities leads to a linear temperature dependence, while the scattering from neutral impurities leads to the usual Fermi-liquid behavior. By considering an appropriate mass action law to determine the ratio of charged to neutral impurities we can describe the experimental metal-insulator transition at low temperatures as a Mott-Hubbard transition. \end{abstract}
\vskip2pc]

\newcommand{\grlo}{\stackrel{>}{<}}
\newcommand{\logr}{\stackrel{<}{>}}

\section{Introduction}

The low temperature conductivity of quasi two-dimensional systems 
like MOSFET structures or hetero-junctions reveals a surprising 
metal to 
insulator transition \cite{AKS01}. This critical review of the theoretical 
approaches has concluded that this phenomenon is insufficiently explained. 
The experimental data  
shows a pronounced transition from insulating 
behavior at low densities to a metallic behavior at 
high densities. 
The generic feature of the metal-insulator transition is the rapid change from
insulating to conducting behavior when the density is increased 
very slightly at low temperatures. This density driven metal-insulator
transitions are usually referred to as Mott transitions \cite{Ge97}. The
characteristic feature of the Mott-Hubbbard transition is that an increase in the effective mass is directly responsible for increasing resistivity
$\rho=m/e^2 n\tau$ while the Anderson scenario would assume a
vanishing relaxation time $n \tau$. Measurements of the 
effective mass \cite{SKDK01} seem to support a Mott-Hubbard transition
rather than the Anderson transition quantitatively explained in Ref. \cite{M02}. 

In this paper we want to return to the original idea of the Mott 
transition in that a bound state is resolved with 
increasing density due to pressure ionization. We will show 
that a quantitative description of the experimental results can 
be achieved if one calculates the interplay of weak localization
and trapping due to charged impurities 
 as well as the scattering with neutral impurities on the same 
theoretical footing. 

Weak localization as a quantum interference effect is intensively
studied in the literature \cite{VW80,AA85,KB86,BK94,J00}. 
The maximally crossed diagrams lead to a
diffusive pole which allows one to extract weak localization corrections
to the conductivity \cite{BK94,aalr,ChS86,DHIKSZ98} (and citations therein). 
This paper is
devoted to an alternative route to weak localization. We will linearize the quantum kinetic equation derived at a lower level of approximation but with external fields in order to create higher order diagrams. The main aim is to show that weak
localization has the same origin as the interference effect known from the Debye
-Onsager relaxation effect in plasma physics.

In the following part of the introduction 
we will outline the model we want to use. This will give a summary about the kinetic approach adopted in this paper. The many-body approximation to be applied in this paper is also specified and it is clarified how higher order diagrams are generated by linearization, which is presented in detail in appendix~\ref{variation}. In the second section we calculate the electrical conductivity from the kinetic equation approach and present results for the relaxation time and  the relaxation function. This relaxation function summarizes the quantum corrections to the conductivity due to interference effects. The Lenard-Balescu kinetic equation used as a starting point in this section
is derived in appendix~\ref{dynscreen}. The third section discusses the resulting conductivity formulae and shows that the relaxation function is identical to
the one calculated for the weak-localization corrections. For comparison with experiment we include scattering from neutral impurities  besides the scattering with charged impurities which are worked out in detail in appendix~\ref{impuritys}.
The fourth section summarizes the results and gives an outlook. In appendix~\ref{pol2D} we discuss the polarization function in quasi two-dimensional systems. The additional appendices~\ref{a1} and \ref{integral} present calculations of the integrals used during the paper.

\subsection{Outline of the model}
As a model, we assume a quasi two-dimensional Coulomb potential, where
the field lines are three dimensional but the motions of
the particles are restricted to two dimensions. The effective
potential takes then the form $V_{ab}=4 \pi e_a e_b \hbar/\sqrt{q_x^2+q_y^2}$
where for instance 
$e_a=e$ for electrons and $e_b=-Z e$ for charged ions.
Our approach is conveniently based on the kinetic
equation for the one-particle distribution $f_a(k,t)$ normalized to the 
area density
\be
s \int { d{\bf k} \over (2 \pi \hbar)^2} f_a({\bf k},t)=n_a
\ee
where the spin degeneracy  is denoted by $s$.
The kinetic equation for that distribution function is
\be
&&\partial_t f_a+e_a {\bf E \nabla_k} f_a=\sum\limits_b{\cal
  I}[f_a,f_b,{\bf E}]
\label{kin}
\ee
where correlations are covered by the corresponding collision integral
${\cal I}$. This collision integral is explicitly field dependent due
to the distortion of two-particle correlations. We consider the
conductivity in a system described by a local equilibrium distribution
\be
f_a({\bf k},t)=\left ({\rm e}^{{({\bf k-p_a}(t))^2-\mu_a\over 2 m_a T_a}}+1\right )^
  {-1}
\label{sh}
\ee
with the mean  mass-motion of the charged particle ${\bf p_a}(t)$. 
The center of
mass motion is at rest, which means that the total sum of currents
$\sum\limits_b {\bf j_b}=\sum\limits_b {n_b |e_b| {\bf p_b}\over m_b}=0$ with
density $n_b$ and mass $m_b$.
In the following we will restrict ourselves to a two-component system. The
generalization towards multicomponent systems is straightforward.

From the collision integral (\ref{kin}) we have two sources of
linear response: A term proportional to the current ${\bf j}=-e_a n_a
{\bf p_a}/m_a$ and a term proportional to the field.
By multiplying (\ref{kin}) with $k$ and integrating, the balance equation for the momentum reads
\be
\partial_t (n_a {\bf p_a})-n_a e_a {\bf E} &=&\int {d k \over (2 \pi
  \hbar)^2} {\bf k}
  {\cal{I}}\nonumber\\&=&-n_a e_a {\bf E} {\delta E \over E}+n_a {\bf p_a} \tau^{-1}
\label{curr}
\ee 
such that the current balance takes the form
\be
\partial_t {\bf j_a}-{n_a e_a^2 \over m_a} \left ( 1-{\delta E\over E}
\right ) {\bf E}=-\tau^{-1} {\bf j_a}
\ee
and the stationary conductivity ${\bf j}=\sigma {\bf E}$ is
\be
\sigma={n_a e_a^2\over m_a} {1-{\delta E\over E}\over \tau^{-1}}.
\label{con}
\ee
According to the distortion of the Fermi function (\ref{sh})
we have a linear response
\be
f_a'-f_a=-{\bf p_a} \left ( \partial_{\bf k_a'}-\partial_{\bf k_a}
\right ) f_0
\ee
and we can 
represent the current relaxation time and the relaxation effect
for a scattering with impurities (\ref{st1}) with the potential $V_s$ as
\be
\left ( \matrix{{1\over \tau_i}\cr {\delta E_i \over E}} \right )&=&-{2 n_i
\over n_a \hbar} \int {d k d q \over (2 \pi \hbar)^4}
\partial_{\epsilon_k} f_0(\epsilon_k) V_s^2(q) q^2 \cos^2(q,E)
\nonumber\\
&&\times \left
  (\matrix{\pi \delta (\epsilon_k-\epsilon_{k-q})\cr {\hbar \over 2}
    {{\cal P}'\over \epsilon_k-\epsilon_{k-q}}} \right ).
\label{transi}
\ee
The relaxation time $\tau$ 
is coming from the term proportional to the 
current in the collision integral. This is due to on-shell scattering represented by the delta function in (\ref{transi}). Besides this the conductivity
becomes renormalized by the explicit field dependence of the collision
integral, ${\delta E\over E}$, which is an interference effect and corresponds to an off-shell scattering as expressed by the derivative of the principal value in (\ref{transi}). 
The latter effect has been the subject of
various investigations for nondegenerate plasmas and is known as
Debye-Onsager relaxation effect. The field
dependence and quantum form is discussed in \cite{Mo00}. For a plasma system
this Debye- Onsager relaxation effect \cite{k58,KE72,e76,er79,r88,MK92,ER98} was first derived within
the theory of electrolytes \cite{DH23,O27,f53,FEK71,KKE66}. \footnote{Debye derived a limiting law of electrical conductivity
\cite{DH23} which stated that the external electric field $E$ on a
single charge $Z=1$ is diminished in an electrolyte solution by
the amount $\delta E/E={\kappa e^2 \over 6 T}$
where  $e$ is the elementary charge, $E$ the electric field strength,  $T$
is the temperature of the plasma and $\kappa$ is the inverse screening
radius of the screening cloud. This law is interpreted as a
deceleration force which is caused by the deformed screening cloud
surrounding the charge. Later it has been shown by Onsager
\cite{O27} that this
result has to be corrected to $\delta E/E={\kappa e^2  \over 3 (2 +\sqrt{2})T}$
if the dynamics of ions ($Z=1$) is considered. 
The linear response theory  reproduce this Onsager result \cite{er79,r88,ER98}.
The kinetic theory leads to the Onsager result if asymmetric screening 
\cite{Mo00} is applied  while the symmetric treatment leads to the Debye result
  \cite{MK92,Ma97,ER98}. }

The theoretical calculations of conductivity in reduced dimensions
is the topic of intensive investigations. 
These concern rigid two dimensional electron systems\cite{Mi87} 
and quasi-two dimensional systems\cite{HSW71}. In the latter 
study a three-dimensional system was considered, where the 
particles can only scatter in two dimensions leading to 
a cylindrical Fermi surface. The Born approximation and contact 
interaction result in a 
resistivity which has a leading low temperature behavior as
${1\over \sigma}\propto a T^2(1+b \ln{T})$.
We will show that the Coulomb interaction with the dynamical 
screening results in a linear order as the leading term. This 
has been repeatedly reported in the literature both from an experimental and
theoretical point of view.
Numerical calculations of Coulomb scattering rates 
from impurities predict a linear temperature dependence of the 
mobility in silicon inversion layers \protect\cite{CW80,SH99}.
This was attributed to the collisional level broadening in the 
screening function. Related results have been obtained in 
Ref.~\cite{S86} where a significant suppression of the temperature 
dependence of the screening function was found. An analytical 
investigation of screening in quasi two-dimensional systems was 
given in Ref.~\cite{GD86} where a linear temperature term in the 
conductivity was reported.

In this paper we want to investigate the effect of Coulomb
screening on the conductivity. We will derive exact analytical
results which show that due to dynamical screening the
 leading low temperature contribution to the conductivity is linear. 
In contrast, the static screening leads to a
quadratic temperature dependence typical for the Fermi liquid.

\subsection{Many body approximation used in this paper}

For the calculation of the conductivity we want to account for quantum 
interference effects  such like weak 
localization. This means we have to include maximally crossed 
diagrams at least \cite{VW80,KB86,DHIKSZ98}. Besides the direct calculation of diagrams, we can employ the philosophy of variational techniques. In Ref. \cite{KB00} it was described how one can use the variation of nonequilibrium Green's functions with respect to an auxiliary external field to create higher order diagrams in the response function. This makes use of the known variation technique summarized in appendix~\ref{variation}.

\begin{figure}
\centerline{\psfig{file=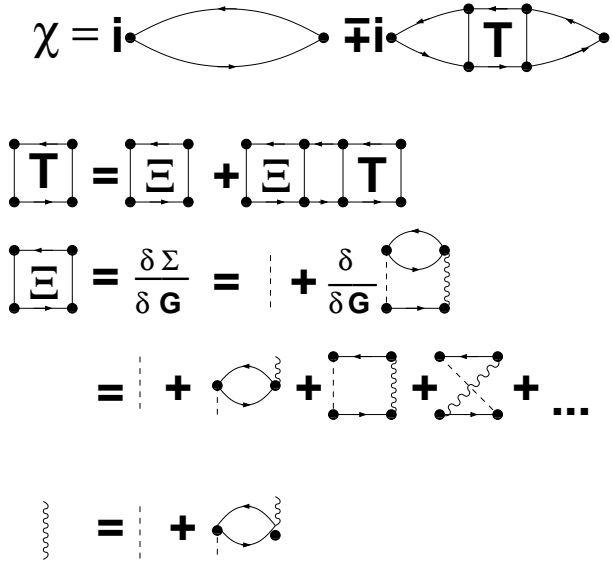,width=8cm}}
\caption{\label{k}The density response function (first line) in terms of
  the particle-hole T-matrix. The latter one can be expressed in the
  second line as a sum of irreducible graphs which are given in terms
  of the selfenergy. The latter one is used in screened potential
  ${\cal V}=V+V \Pi{\cal V}$ (last line) approximation.}
\end{figure}

The response function can be given in terms of the particle hole
T-matrix which in turn can be represented by the sum of irreducible
graphs, $\Xi$,
according to figure \ref{k}. In turn, $\Xi$ can be expressed
as a variation of the selfenergy $\Sigma$ with respect to the
Green's function (bare line) $G$. For the dynamical screened
approximation used here 
we give the corresponding results in figure \ref{k}. One
sees that in principle maximally crossed diagrams are accounted
for. 

Instead of the variation of the nonequilibrium Green's function we can use a proper reduction of the latter ones towards a kinetic equation.
The linear response obtained from this
kinetic equation including all external field effects is accounting then for
higher order diagrams in a convenient way. 
As such, we obtain weak localization effects by proper linearization of
the collision integral.

\section{Kinetic equation and conductivity}
The kinetic equation corresponding to the dynamical screened
approximation 
is the quantum Lenard - Balescu equation, which has been derived for
high fields in \cite{Mo00,Moa93}. A sketch of the derivation is in
appendix~\ref{dynscreen}
\beq
&&\frac{ \partial}{\partial t}  f_a + e {\bf E} {\pa {\bf k_a}} f_a
=I^{\rm in}_a({\bf k},t)-I^{\rm out}_a({\bf k},t).
\label{kin1}
\ee
The collision-in integral is [$I^{\rm out}$ is given by
$f\leftrightarrow 1-f$ and  $L^< \leftrightarrow L^>$]
\beq
&&I^{\rm in}_a({\bf k},t)=2 \sum\limits_b s_b \int {d {\bf q}\over
  (2\pi\hbar)^2} V_{ab}^2({\bf q}) \int\limits_{0}^{\infty}d \tau \int
{d \omega \over 2 \pi}
\nonumber\\
&&\times \cos \left [(\epsilon^a_{k-q}-\epsilon_k^a-\hbar \omega)
  {\tau \over \hbar}+{e_a {\bf E q} \tau^2 \over 2 m_a \hbar} \right ]\nonumber\\
&&\times
f_a({\bf k\!-\!q}\!-\!e_a {\bf E} \tau,t\!-\!\tau) (1\!-\!f_a({\bf k}\!-\!e_a {\bf
  E}\tau,t\!-\!\tau))
\nonumber\\
&&\times
{\Pi^<_{bb}({\bf q},\omega, t-\frac 1 2 \tau)\over \left | {\cal E}({\bf
      q},\omega,t-\frac 1 2 \tau) \right |^2} 
\label{eqe}
\eeq
with the free density fluctuation (\ref{fluc})
\beq\label{fluc1}
&&\Pi_{bb}^<({\bf q},\omega,t)=-2 \int {d {\bf p}\over (2\pi\hbar)^2}
\int\limits_{0}^{\infty} d\tau 
\nonumber\\
&&\times
\cos \left [ (\hbar \omega-\epsilon_p^b+\epsilon_{p+q}^b){\tau \over \hbar} +{e_b {\bf E
      q} \tau^2 \over 2 m_b \hbar}\right ] 
\nonumber\\
&&\times
f_b({\bf p+q},t-\frac 1 2 \tau) (1-f_b({\bf p},t-\frac 1 2 \tau)).
\eeq
Here we use the sum over species explicitly.

The nested form of the Lenard - Balescu collision integral (\ref{eqe})
is computationally advantageous. It tells that the collision
of particle $a$ with momentum $k$ on a particle $b$ with momentum $p$
into $(a, k-q)$ and $(b,p+q)$ can be represented equivalently as a collision of the particle
$a$ with $k$ on a hole $a$ with $k-q$ by a dynamic plasmon emission
which is considered as a particle-hole fluctuation of particles $b$.
For static screening ${\cal E}({\bf q},0,t)$, equation
(\ref{eqe}) reduces to the kinetic equation for statically screened
Coulomb potentials in high
electric fields (\ref{kinetic})
\protect\cite{JW84,Moa93}.

Now we 
calculate the frequency integral in (\ref{kin1}) and (\ref{eqe}) analytically 
using the
identity (\ref{a13}).  We summarize the result
of the frequency integration in the momentum dependent function
$W(q)$ which takes the explicit form (\ref{W}) for quasi
two-dimensional systems.

Performing the balance equation for the current (\ref{curr}) we obtain
in linear response the interference term or relaxation function
\be
e_a n_a {\delta E\over E}&=&{2 e_a\over m_a \hbar^4}\sum\limits_b s_b \int {d q \over (2
  \pi \hbar )^2} W(q) q^2 \cos^2(q,E) \nonumber\\&&
\times\int\limits_0^\infty d \tau
\tau^2 I_s(a,\tau) (-\frac 1 2 \Pi^<_{bb}(q,0))
-(a\leftrightarrow b)\nonumber\\
\label{relaxf}
\ee
and the relaxation time as
\be
\tau^{-1}=\sum\limits_b \left (R(a,b)-{e_a m_b \over e_b m_a}
  R(b,a)\right )
\ee
with
\be
  n_a R(a,b)&=&-{4 s_b \over m_a \hbar^4} \int {d q \over (2
  \pi \hbar )^2} W(q) q^2 \cos^2(q,E) \nonumber\\&&
\times\int\limits_0^\infty d \tau
\tau  I_s(a,\tau) (-\frac 1 2 \Pi^<_{bb}(q,0)).
\nonumber\\&&
\label{rf}
\ee
Here the static free density fluctuation reads
\be                                                     
\Pi^>_{bb}(q,0)&=&\Pi_{bb}^<(q,0)\nonumber\\&=&2 \pi \int {d q \over (2 \pi \hbar)^2}
  f_p^b(1-f_p^b) \delta(\epsilon_p-\epsilon_{p+q})          
\nonumber\\&&                                           
\label{po}
\ee                                                     
and the integral
\be
I_s(a,\tau)&=&\int {d k\over (2 \pi\hbar)^2} f_a(k)(1-f_a(k-q)) 
\nonumber\\ &&\times \sin{\left (
    \epsilon_{k-q}-\epsilon_k \right )
{\tau\over \hbar}}
\label{isa}
\ee
will be calculated in appendix~\ref{integral}.

We remark that the interference effect (\ref{relaxf}) vanishes for identical
scattering partners, e.g. electron-electron correlations. 

When calculating the explicit form of the relaxation time and
relaxation function we employ charge neutrality $e_a n_a+e_b n_b=0$
and restrict ourselves to the case of single charge ions $e_b=-e$. The
case of higher charged ions is also available but is more involved. The 
generalization to systems with additional particles species 
is straight forward.

\subsubsection{Relaxation time by charged impurities}

Introducing the dimensionless integration variable $q=2 p_{fa} y$ we obtain
the relaxation time to lowest order in temperature
\be
\tau^{-1}&=&\sum\limits_b{m_a e_a^2 p_{fa} s_b \over \pi \hbar^4 n_a}
\kappa_a  \left (T_b- {e_a n_a
    T_a \over e_b n_b}\right ) \nonumber\\&&\times{\xi\over \xi+1} \int\limits_0^{1} d y
{y \kappa_a\over
  (y+\kappa_a)}{1\over \sqrt{1-y^2}}
\label{t}
\ee
where we have introduced the abbreviation
$\xi={m_b^2 e_b^2\over m_a^2 e_a^2}$.
Since the momentum integration $y \le 1$ is restricted by the low
temperature expansions to values below $2 p_{fa}$ the low temperature
expansion of the inverse screening length (\ref{kq}) becomes a constant
\beq\label{screen1}
\kappa&=&\sum\limits_c {2 \pi e_c^2 \partial_\mu n_c}
\nonumber\\
&=&{e_a^2 m_a s_a\over \hbar ^2} \left ( 1+ \left |{e_b s_b\over
      e_a s_a}\right | \sqrt{\xi}
\right ).
\eeq
We distinguish here between the temperature of electrons $T_a$ and the temperature of the ions $T_b$ which could mime nonequilibrium effects.

The integral in (\ref{t}) can be easily calculated
\be
R=\tau_{ab}^{-1}&& {m_a\over n_a e_a^2}{e_a^2\over h}= {8 s_b\over s_a^3 } \left ({T_a\over \epsilon_{fa}}+{T_b\over
    \epsilon_{fa}}\right ) {\xi\over 1+\xi} \kappa_a' \kappa_a \nonumber\\
&&\times\left (
{\pi \over 2} +{\kappa_a'\over
      \sqrt{1-\kappa_a'^2}} \ln{{\kappa_a'
\over(1+\sqrt{1-\kappa_a'^2})}}
\right )
\ee 
where we will use the abbreviation
\be 
\kappa_a={\hbar \kappa\over 2p_{fa}}={e_a^2 m_a s_a\over 2 \hbar
  p_{fa}}=\sqrt{s_a^{3} e_a^4 m_a^2\over 16 \pi \hbar^4 n_a}
\equiv \sqrt{\hbar \over \tau_0 \epsilon_f}
\label{t0}
\ee
with$\tau_0^{-1}=m e^4 s^2/8/\hbar^3$ during the paper.
Furthermore, we distinguish in (\ref{t0}) between the inverse screening length where ions are included, $\kappa_a'$, and where they are neglected $\kappa_a$ 
\be
\kappa_a'&=&\kappa_a (1+\sqrt{\xi} {s_b\over s_a}).
\label{58}
\ee

We expand the above result
for large and small $\kappa_a$ which corresponds also to the small and
large density limits
\be
R&=& {8 s_b\over s_a^3 } \left ({T_a\over \epsilon_{fa}}+{T_b\over
    \epsilon_{fa}}\right ) {\xi\over 1+\xi} \kappa_a 
\nonumber\\&&\times
\left \{
\matrix{
1-{\pi\over 4 \kappa_a'}+{2 \over 3 \kappa_a'^2}+o(\kappa_a'^{-3})
\cr\cr
{\pi\over 2}\kappa_a+\kappa_a^2\ln{\kappa_a\over 2}+o(\kappa_a^{3})
}
\right ..
\label{tl}
\ee

It is interesting to investigate the limit of large ion masses, $\xi\to\infty$
and $T_b=0$, which would correspond to 
the charged impurity limit.  One gets from (\ref{tl}) for large $\kappa_a$
\be
R&=& {8  s_b\over s_a^3 } {T_a\over \epsilon_{fa}} \kappa_a .
\label{tli}
\ee
If we compare this with the neutral electron-impurity scattering result (\ref{til}), we see significant
differences. While the statically screened result shows a Fermi-liquid
behavior of $const$ and $T^2$ terms, the dynamically screened result
(\ref{tli}) leads to a linear temperature dependence.

\subsubsection{Relaxation function by charged impurities}

Interference effects from the relaxation function (\ref{relaxf}) 
can be calculated analogously
\be
&&{\delta E\over E}={e_a m_a^2 s_b \over 2 \pi p_{fa}  \hbar^3
  n_a}\left (e_b T_a -e_a T_b \right ) {\xi\over \xi+1} 
\nonumber\\&&\times \int\limits_0^{\infty} d y
{\kappa_a'\over
 y (y+\kappa_a')}\partial_y^2\sqrt{y^2-1} \Theta(y-1\pm\eta)
\label{t1}
\ee
The small $\eta$ has been introduced to perform the principal value
integration ${\cal P}$ according to (\ref{istt}). It should be noted
that the $\Theta$ functions of the denominator and numerator cancel
exactly and no restriction on $y$-integration remains.
 
We have now to carefully consider the structure
\be
&&{\cal I}_\pm=\int\limits_0^ \infty dy f_y {\partial^2 \over
  \partial y^2} \left [g_y \Theta(y-1\pm\eta) \right ]=
\nonumber\\
&&\times \lim\limits_{\eta \to 0} \left (
  \int\limits_{1\mp\eta}^{\infty} dy f_y g_y''+(f_y g_y'-f_y'g_y)|_{y=1\mp\eta}
\right )
\label{struc}
\ee
with $f_y=\kappa_a'/y/(y+\kappa_a)$ and $g_y=\sqrt{y^2-1}$. 
Performing the integral one sees that
the divergent contribution at $\eta\to 0$ is cancelled exactly by
the $f g'-f'g $ term. We obtain
\be
{\cal I}_\pm= {\kappa_a'\over 1-\kappa_a'^2}\pm {\pi\over 2}+{{\rm arcosh}
  \kappa_a'\over (\kappa_a'^2-1)^{3/2}}.
\ee
The principal value in (\ref{t}) is calculated from $({\cal I}_++{\cal
  I}_-)/2$
and we obtain (with charge $e_b=-e_a$)
\be
{\delta E\over E}&=&-{2 s_b\over s_a^2} \left ({T_a\over \epsilon_{fa}}+{T_b\over
    \epsilon_{fa}} \right ) \kappa_a {\xi\over 1+\xi}
\nonumber\\&&\times \left ({\kappa_a'\over 1-\kappa_a'^2}+{{\rm arcosh}
  \kappa_a'\over (\kappa_a'^2-1)^{3/2}}\right ).\nonumber\\&&
\label{e}
\ee
The low density (large $\kappa_a$) expansion as well as the high
density (small $\kappa_a$) expansion read \footnote{It is clear that for the impurity limit with infinite masses
$\xi\to\infty$ we have from (\ref{58}) and (\ref{el})
\be
\left ({\delta E\over E}\right )_i&=&0
\ee
in agreement with the physical picture that if the ions cannot move
the screening cloud cannot deform during the motion of the electrons and
cannot induce a relaxation effect. This is different if the charged
impurities do not contribute to the screening,
$\kappa_a=\kappa_a'$ in (\ref{58}), and we obtain a finite result. This case is
anticipated here since the neutral static relaxation function
already lead
to finite results (\ref{imprel1}).}
\be
{\delta E\over E}&=&-{2 s_b\over s_a^2} \left ({T_a\over \epsilon_{fa}}+{T_b\over
    \epsilon_{fa}} \right ) \kappa_a {\xi\over 1+\xi}
\nonumber\\&&\times \left \{
\matrix{
-{1\over \kappa_a'}+\left(\ln{(2\kappa_a')}-1\right ){1\over
  \kappa_a'^3}+o(\kappa_a'^{-4})
\cr\cr
-{\pi\over 2}+2 \kappa_a'-{3 \pi\over 4} \kappa_a'^2 +o(\kappa_a'^{3})
}
\right ..\nonumber\\&&
\label{el}
\ee

\section{Discussion}

For further progress we use the expansions for
large $\kappa_a$ or low densities.
Collecting equations (\ref{tl}) and (\ref{el}) we obtain the
conductivity in an analogous form to the Bloch-Gr\"uneisen formulae

\be
\sigma={e^2\over h} {1-{\delta E\over E
    }(n)\over {R(n)} }
\label{bloch}
\ee
where the dynamic parts comes from the scattering from charged
impurities (\ref{tl}), (\ref{el}) 
\be
R &=& {8 s_b\over s_a^3 } \left ({T_a\over \epsilon_{fa}}+{T_b\over
    \epsilon_{fa}}\right ) \kappa_a
\left (1-{\pi\over 4 \kappa_a}+{2 \over 3
    \kappa_a^2}+o(\kappa_a^{-3})\right )
\nonumber\\
{\delta E\over E}&=&{2 s_b\over s_a^2} \left ({T_a\over \epsilon_{fa}}\!+\!{T_b\over
    \epsilon_{fa}} \right ) 
\left (
{1}\!-\!\left(\ln{(2\kappa_a)}\!-\!1\right ){1\over \kappa_a^2}\!+\!o(\kappa_a^{-4})
\right ).
\nonumber\\&&
\label{bloch3}
\ee
We find that both the relaxation time as well as the relaxation function have a linear temperature dependence for temperatures lower than the Fermi energy. This is in agreement with the experimental and theoretical works mentioned in the introduction.

\subsection{Relation to weak localization}

The low density or weak potential limit (\ref{bloch}) is now
interesting to discuss. Using (\ref{t0}) we can rewrite (\ref{bloch})
into a net relaxation effect [$s_a=s_b=2$, $T_a=T_b$]
\be
{\sigma\over \sigma_0}&=&
1-{{ \delta
      E}\over { E}}
\nonumber\\&=&
1-2{T\over \epsilon_f} \left (1+{\tau_0
      \epsilon_f \over 2 \hbar} \ln{\left( {\tau_0 \epsilon_f\over 4
      \hbar}\right )} \right )
\nonumber\\&=&
1
-2 {T\over \epsilon_f}-{\hbar \over \tau
      \epsilon_f} \ln{\left( {\hbar \over 4 \tau T}\right )}
\label{anderson}
\ee
where we have introduced the temperature dependent relaxation time 
$1/\tau=\epsilon_f T \tau_0/\hbar^2= 8
\epsilon_f \hbar T/m e^4 s^2$.
This is precisely the localization correction to the
conductivity \cite{KB86,AA85}. 
Therefore we understand now the physical meaning of the relaxation
function at low temperatures. Please note that it vanishes here
for small temperatures in contrast to the static result (\ref{imprel1}).

\subsection{Comparison with experiments}

In order to describe realistic experiments, we extend our model with scattering from
static neutral impurities (treated in
appendix~\ref{imprel1}).
In addition to (\ref{bloch}) we have static contributions coming 
from the scattering from neutral
impurities (\ref{til}),(\ref{imprel1}) as
\be
R_i&=&{2^{3/2}s_i n_i\over
  s_a n_a}\left ({a_0\over 2 r_0}\right )^2 
\left
  \{
{1}\!-\!{\pi\over 2 \kappa_i}
\!-\!{\pi^2\over 24}{T^2  \over
    e_{fa}^2}{\pi^3\over 16 \kappa_i}
\!+\!o(\kappa_i^{-4})
\right \} \nonumber\\
{\delta E_i\over \delta E}&=&{n_i\over
  n_a}\left ({a_0\over 2 r_0}\right )^2 
\left
  \{
{1}+ {4-3\ln{2 \kappa_i} \over \kappa_i^2}
+o(\kappa_i^{-4})
\right \} 
\label{bloch2}
\ee

We can safely use the large $\kappa_a$ limit since typical densities of
\cite{KK00} are $7\times 10^{10}$cm$^{-2}$ and we have
\be
\kappa_a={284.9\over \sqrt{n_a/7\times 10^{-10}{\rm cm}^{-2}}}
{m^*\over m}
\ee
which is a large parameter.

\begin{figure}
\psfig{file=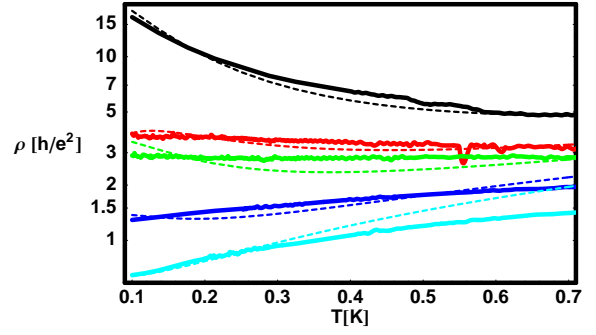,width=8cm}
\caption{The conductivity versus temperature according to
  (\protect\ref{bloch1}) as dashed lines. The experimental curves are the solid lines \cite{KK00}. From top to bottom they correspond to densities of 
$n_a=6.85,7.17,7.25,7.57,7.85 \times 10^{10}$cm$^{-2}$.  \label{mos_fit}}
\end{figure}

Now we have two unknown fit parameters in the theory. These are the energy
level $E_b=E_c-E_D$ of the impurities determining the ratio of
impurity density to electron densities \cite{IL96,M02}
\be
{n_i\over n_a}=n_a \left ({2 \pi \hbar^2\over m_a T}\right ) {\rm e}^{\beta E_b}
\label{mott}
\ee
 and the ratio of the scattering
length to the scattering range of neutral impurities $\zeta=\left (a_0\over r_0 \right )^2 \epsilon_{fa}$. 
From (\ref{bloch}) we can write the final conductivity formula in the
form
\be
{\sigma\over \sigma_0}&=&1-{\zeta\over 2 T} {\rm e}^{E_b/T}-{T\over 2
  \epsilon_{fa}}+{T\over 2 \epsilon_{fa} \kappa_a^2} \ln{2
  \kappa_a\over e}\nonumber\\
\sigma_0&=&{e^2\over h} \left ( {\zeta\over \sqrt{2} T} {\rm e}^{E_b/T}
+{T\over \epsilon_{fa}} \left (\kappa_a-\frac \pi 4 \right ) \right )
\label{bloch1}
\ee
where we have used the spin degeneracy of the heavy impurities $s_b=1$ and
the temperature $T_b=0$. 
The best fits to the experimental results \cite{KK00} are plotted in figure
\ref{mos_fit}.

We see in figure \ref{mos_fit} a clear insulator to metal transition for
low temperatures when the density is increased very slightly. The fitting formula (\ref{bloch1}) works quite well
at all experimental densities for low temperatures, however the formula fails
for higher
temperatures. This is because we used the low
temperature Sommerfeld expansion and the Fermi energy is  
$1.9$ K $\times n_a/7\times 10^{10} cm^{-2}$ in this case, such that at $0.8$K we expect deviations from the leading low temperature behavior. Despite this
imperfect agreement with the data, it is quite satisfying that the metal to insulating transition can be described completely by the scattering with charged and neutral donor impurities supplemented by a mass action law. This strongly favors the Mott-Hubbard transition picture. 

\begin{figure}
\psfig{file=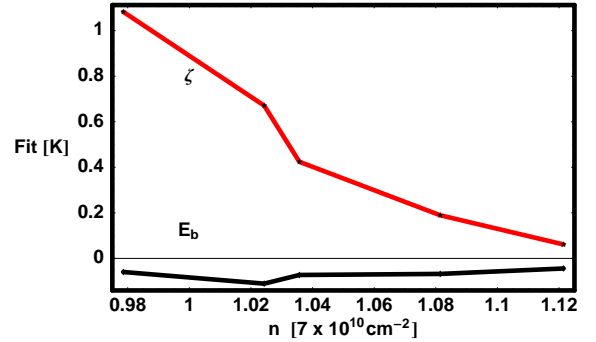,width=8cm}
\caption{The fit parameter $E_b$ and $\zeta$ from
  (\protect\ref{bloch1}) versus density. 
\label{mos_fit_par}}
\end{figure}

The best fit parameter are shown
in figure~\ref{mos_fit_par}. 
We see that the effective binding energy of the electrons to the impurities as well as the scattering strength decrease rapidly when passing through the critical density. This is in agreement with the picture that we have pressure ionization i.e. a crossing between continuum and bound state levels. Therefore, this underlines the Hubbard-Mott transition which we have seen already in the increase of the effective mass \cite{SKDK01} as described in \cite{M02}.

A remark concerning the formula (\ref{bloch1}) should be made here. Instead of
the trapping at charged impurities resulting in the  mass action law (\ref{mott}), one could assume in principle any other trapping mechanism. In particular, we have shown that three-particle bound states can describe the experimental data as well \cite{NM01}. The underlying conductivity formula is precisely (\ref{bloch1}) where the number of three-particle bound states diminishes the charged impurity density correspondingly. So the basic transport mechanisms outlined in this paper here remain the same. The only mechanism that is not possible to extract so far is that of actual trapping.

\section{Summary and Outlook}

Linearizing the Lenard - Balescu collision integral including all external field dependences allows one to derive a conductivity on the level of an infinite series of diagrams including maximally crossed lines. The field dependence of the collision integral yields an interference effect which is shown to describe just the weak localization corrections. This has the same formal origin as the Debye-Onsager relaxation effect in plasma and electrolyte systems.

For the low temperature regime, it was possible to calculate the conductivity analytically. It was found that the dynamical screening by charged impurities leads to a linear temperature dependence of the conductivity, while neutral impurities give rise to the usual Fermi liquid behavior. This finding is general for any scattering of light particles from heavy particles. Therefore it might also be of use for scattering rates in high $T_c$ cuprates \cite{H02}.

The comparison with experiment is performed assuming an appropriate mass action law between the charged donor impurities and the neutral ones which are considered to be captured electrons. The experimental metal-insulator transition can be described quantitatively by fitting the effective binding energy and the unknown scattering strength.

As noted, a similar quality of description of the experimental data was achieved assuming a three particle clustering instead of trapping of electrons. The latter process, however,  relies on the same transport picture as outlined in this paper. Only the composition of neutral and charged impurities is determined differently. So far we cannot determine which process is actually happening. In order to achieve this we must study the magnetic field dependence, which shows quite unique and remarkable features in the experiment. This is left for further work.

To conclude, we suggest that the metal insulator transition found in experiments can be described within a Mott-Hubbard transition scenario in agreement with the effective mass measurements \cite{M02}.

\acknowledgements

Many discussions with Peter Fulde, James Peter Hague, Rajesh Narayanan, Debanand Sa and Nicolas Shannon are gratefully acknowledged.
To Enver Nakhmedov I am especially indebted for suggesting to me the problem
of metal to insulator transition.

\appendix

\section{Variational technique of linear response}\label{variation}

Assuming besides the interaction potential $V_{11'}$ also a coupling of an external potential $U_{11'}$ where the numbers sign cumulative indices like space time,... coordinates, we can express the two particle Green function $G_{121'2'}=1/i^2 <Ta_1a_2 a_2^+a_1^+>$ by a variation of the one particle Green function $G_{12}=1/i<Ta_1a_2^+>$ with respect to the external potential \cite{KB62,kker86} as
\be
G_{121'2'}=G_{11'}G_{22'}\mp {\delta G_{11'}\over \delta U_{2'2}}
\label{1}
\ee
where the upper sign denotes the Fermi and the lower the Bose functions.
Using the Dyson equation
\be
G^{-1}=G_0^{-1}-\Sigma-U
\ee 
we can calculate the derivative in (\ref{1}) and
with the help of the chain rule and $\delta G=-G \delta G^{-1} G$, 
one can express the fluctuation function as
\be
L_{121'2'}&=&G_{121'2'}-G_{11'}G_{22'}\nonumber\\
&=&\mp G_{12'}G_{21'}\mp G_{13}{\delta \Sigma_{34}\over \delta U_{2'2}} G_{41'}
\nonumber\\
&=&\mp G_{12'}G_{21'}+G_{13} {\delta \Sigma_{34}\over \delta G_{56}} L_{5262'}G_{41'}.
\label{3}
\ee
Double occurring indices are understood as integrated over.
With the definition of the occurring vertex function we can express this graphically 
\begin{figure}
\psfig{file=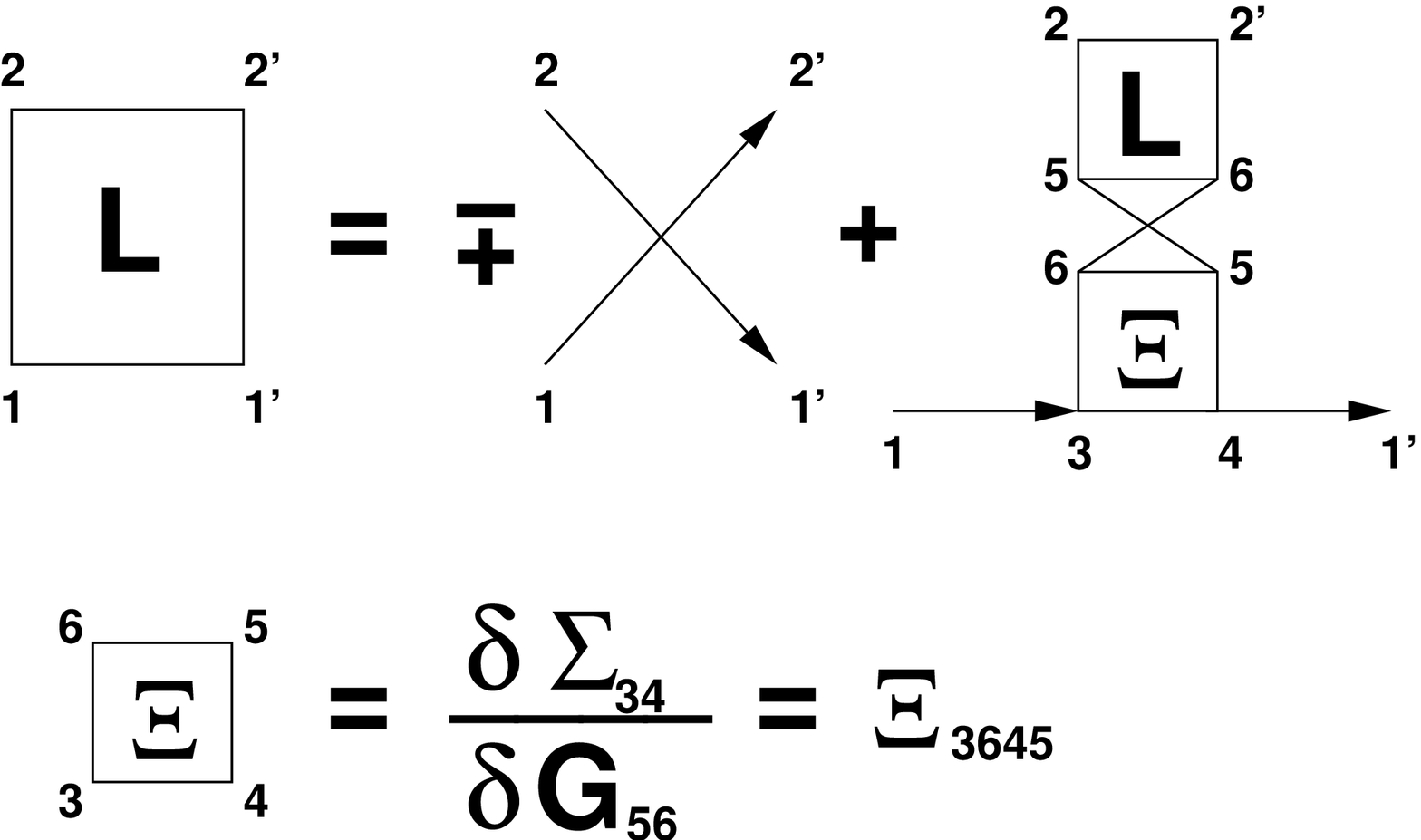,width=8cm}
\caption{\label{v1}}\end{figure}\n
Sometimes it is of advantage to express this density fluctuation function by the T-matrix. Defining
\begin{figure}
\psfig{file=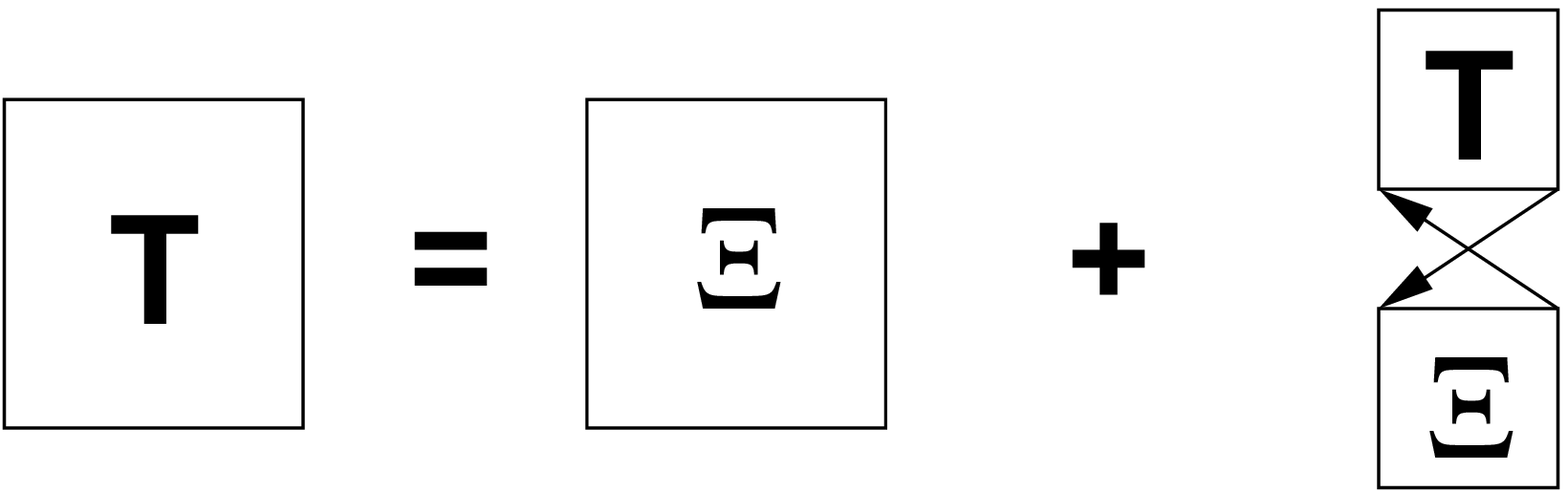,width=7cm}
\caption{\label{v20}}\end{figure}\n 
we can express
\begin{figure}
\psfig{file=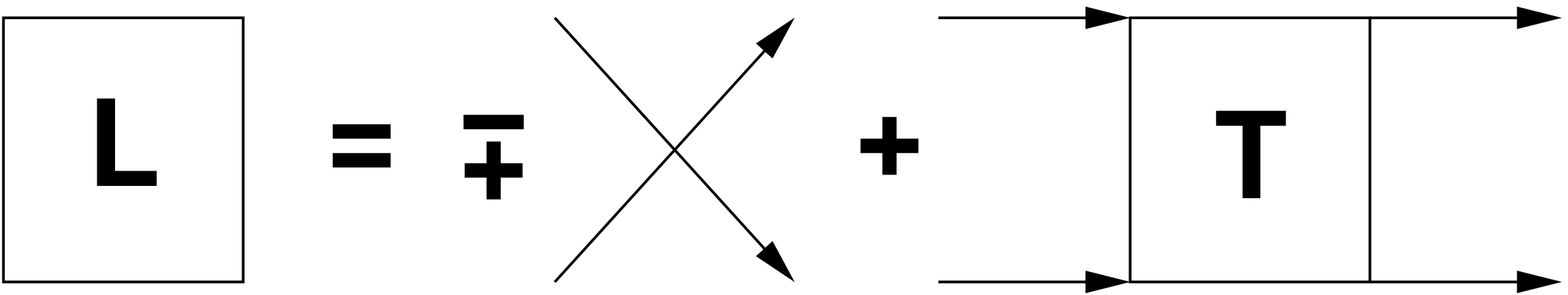,width=7cm}
\caption{\label{v21}}\end{figure}\n 

With the help of (\ref{1}), the density response to an external potential can be expressed in terms of the density fluctuation function $L$ of (\ref{3}). Therefore we remark that the density is given by $i G_{11^+}=<a_1^+a_1>=<\hat n_1>=n_1$ and we have from (\ref{1}) and (\ref{3}) for the response function $\chi$
\be
\chi_{12}={\delta n_1\over \delta U_{22}}=\mp i L_{121^+2}=\pm i <(\hat n_1-n_1)(\hat n_2-n_2)>.
\ee
The last identity follows from the definition of $L$ and underlines the names density fluctuation function.
We see now that the linear density variation due to an external potential can be expressed as
\be
{1\over i} n_1^1=\mp L_{121^+2^+}^0 U_2={1\over i}\chi_{12} U_{22}
\label{15}
\ee
where the upper index indicates the order of external field dependence.
Graphically we can express it as
\begin{figure}
\psfig{file=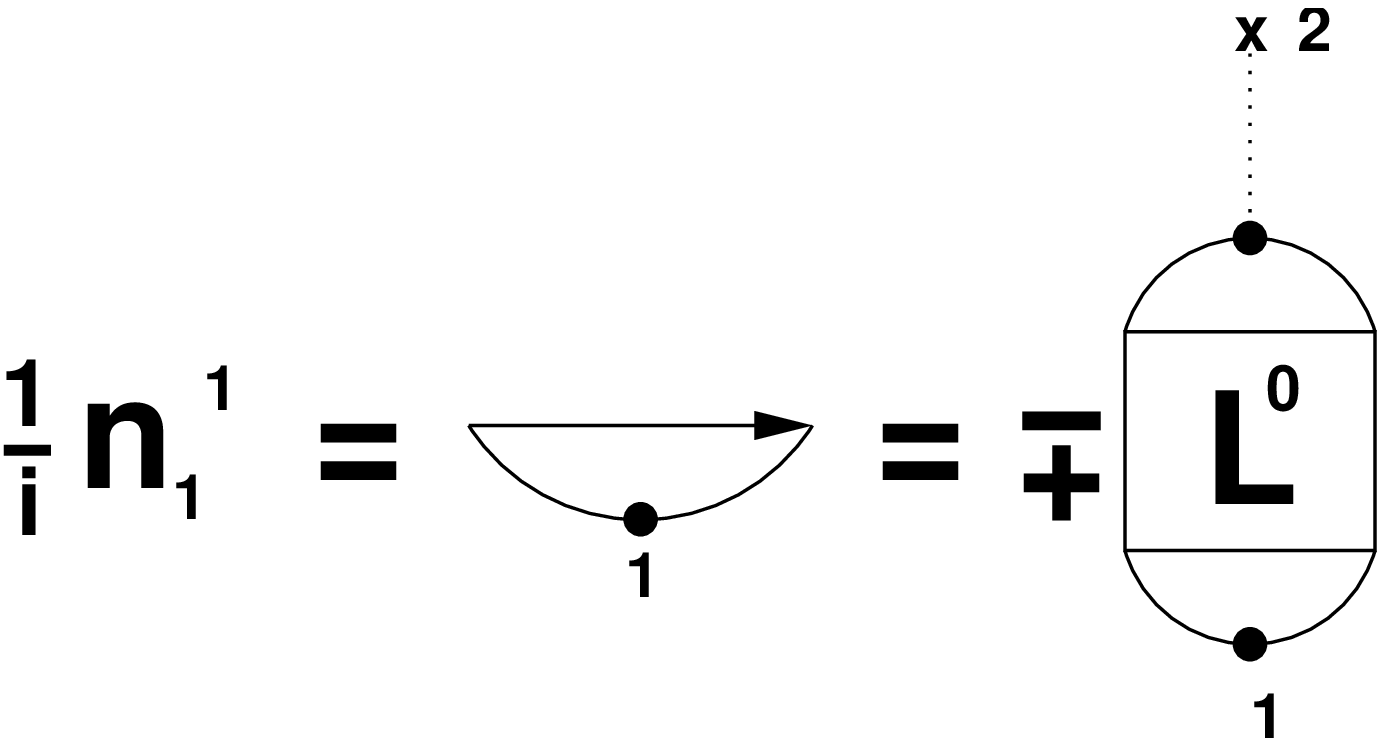,width=7cm}
\caption{\label{v12}}\end{figure}\n
where we will design the external field as a dotted line ending with a cross.
According to (\ref{15}) and figure \ref{v12} we can express the first order response function as
\begin{figure}
\psfig{file=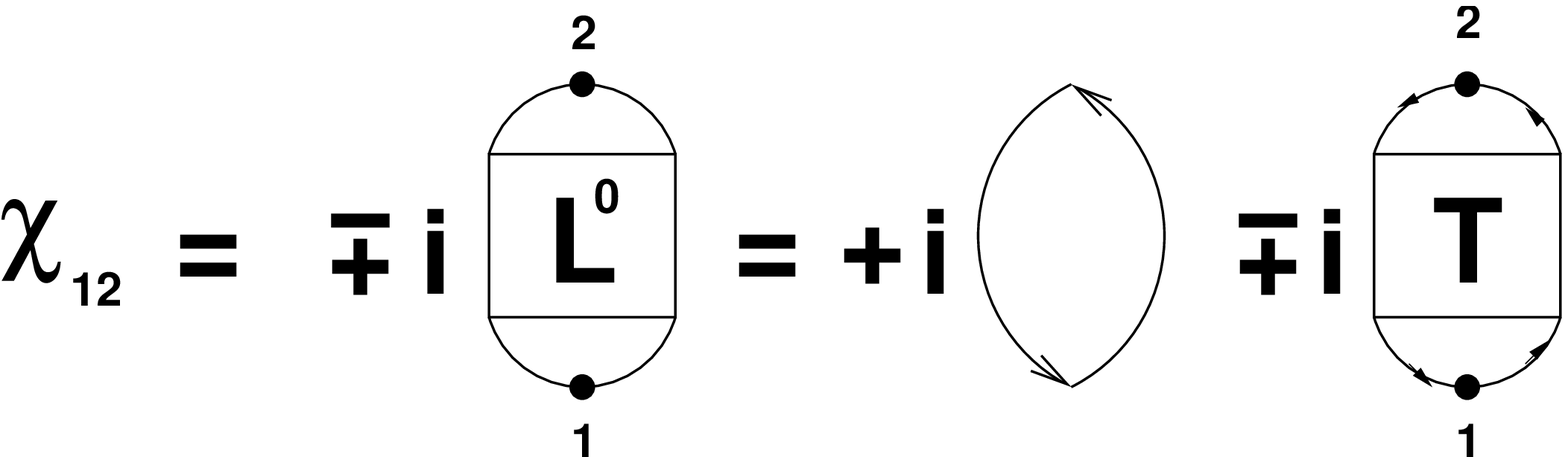,width=8cm}
\caption{\label{v18}}\end{figure}\n
where we have used the definition of the T-matrix in the u-channel figure \ref{v20}. The figures (\ref{v1}), (\ref{v20}) and (\ref{v18}) constitutes the approximation used in this paper (\ref{k}).

\section{Dynamical screened approximation in nonequilibrium}\label{dynscreen}

The selfenergy is given in terms of the dynamical potential ${\cal V}$ according to figure~\ref{k}
\beq
&&\Sigma^<_a({\bf k},t,t')=\int {d {\bf q}\over (2\pi\hbar)^2} {\cal V}^<_{aa}({\bf q},t,t') G_a^<({\bf k-q},t,t')\label{53}
\nonumber\\
&&
\eeq
where the dynamical potential is expressed within Coulomb potentials $V_{ab}({\bf q})$
\beq
{\cal V}^<_{aa}({\bf q},t,t')=\sum\limits_{dc}V_{ad}({\bf q}) {\cal L}^<_{dc}({\bf q},t,t') V_{ca}({\bf q})
\eeq
via the density-density fluctuation
\beq
&&{\cal L}^<_{ab}({\bf q},t,t')=\delta_{ab} \int d{\bar t} d{\bar {\bar t}}\nonumber\\
&&\times
\left ({\cal E}^{r} \right)^{-1} ({\bf q},t,{\bar t})  \Pi_{aa}^<({\bf q},{\bar t},{\bar {\bar t}})
\left ({\cal E}^{a}\right )^{-1}({\bf q},{\bar {\bar t}},t').\label{densityf}
\eeq
Here $\Pi$ is the free density fluctuation or polarization function
\beq
&&\Pi^<_{aa}({\bf q},t,t')=\int {d {\bf p}\over (2\pi\hbar)^2}G_a^<({\bf p},t,t') G_a^>({\bf p-q},t',t)\label{fluc}
\nonumber\\&&
\eeq
and ${\cal E}^{r/a}$ the retarded/ advanced dielectric function
\beq
&&{\cal E}^{r/a}({\bf q},t,t')=\delta(t-t')\pm i \Theta[\pm(t-t')] \sum\limits_b V_{bb}({\bf q})\nonumber\\
&& \times(\Pi^>({\bf q},t,t')-\Pi^<({\bf q},t,t')).\label{57}
\eeq
One easily convince oneself that this set of equations (\ref{53}-\ref{57}) is
gauge invariant. The correlation or Green's function can be related to
the Wigner distribution $f_a$ by the GKB ansatz \cite{LSV86}
\be\label{Lipavsky}
&&G^<({\bf k},\tau,t)=\exp{\left \{-\frac{i}{\hbar}
    \left(\epsilon_k \tau + \frac{e^2E^2}{24m}\tau^3\right)\right \}}
\nonumber\\
&&\times
f \left( {\bf k}-\frac{e{\bf E}|\tau|}{2},t-\frac{|\tau|}{2} \right)
\ee
and analogously for $G^>$ by replacing $f\leftrightarrow (1-f)$.

With the help of the gauge invariant formulation of Green's function, 
we can write the kinetic equation
for the Wigner function $f({\bf p},t)=G^{<}({\bf p,R},t,\tau=0)$
finally \protect\cite{JW84}
\begin{eqnarray}\label{tdinv}
&&\frac{\partial}{\partial t}
f({\bf k},t)+e{\bf E}\nabla_{\bf k} f({\bf k},t)
=\int\limits_0^{t-t_0} d\tau
\nonumber\\
&&\left[\!
\left \{\! G^>({\bf k}\!-\!\frac{e{\bf E}}{2}\tau,\tau,t\!-\!\frac{\tau}{2}),
\Sigma^<({\bf k}\!-\!\frac{e{\bf
    E}}{2}\tau,-\tau,t\!-\!\frac{\tau}{2}) \!\right \}_+\right .\nonumber\\
&&\!-\!\left .
\left\{ \! G^<({\bf k}\!-\!\frac{e{\bf E}}{2}\tau,\tau,t\!-\!\frac{\tau}{2}),
\Sigma^>({\bf k}\!-\!\frac{e{\bf E}}{2}\tau,-\tau,t\!-\!\frac{\tau}{2}) \right \}_+ \right]. \nonumber\\
&&
\end{eqnarray}
This kinetic equation is exact in time convolutions. This is necessary because
gradient expansions in time are connected with linearization in electric fields and consequently
fail \protect\cite{m87}. The gradient approximation in space has been
applied assuming slow varying processes in space  and we have dropped all R-dependence for
simplicity. 
Introducing (\ref{53}) into the equation for
the Wigner function (\ref{tdinv}) one obtains the kinetic equation
(\ref{kin1}) with the explicit form of collision integral (\ref{eqe}).

\section{Statically screened or finite range impurity scattering}
\label{impuritys}

Using the static approximation for the dielectric function ${\cal
E}({\bf q},0,t)$
in (\ref{eqe}), the kinetic equation for statically screened
Coulomb potentials in high
electric fields  appears
\protect\cite{JW84,HJ96}
\begin{eqnarray}\label{kinetic}
&&\frac{ \partial}{\partial t}  f_a + e {\bf E} {\pa {\bf k_a}} f_a =
\sum_b  I_{ab} \nonumber \\
&&I_{ab} =
\!\frac{2 s_b}{\hbar^2}\!\int \!\! \frac{ d {\bf k'_a} d {\bf k_b} d {\bf k'_b}
}{(2\pi\hbar)^4}
\delta \left({\bf k_a}\!+\!{\bf k_b}\!-\! {\bf k'_a}\!-\!{\bf k'_b}
\right)
\nonumber\\
&&\times V_{s}^2({\bf k_a}\!-\! {\bf k'_a},t)
\int\limits_0^{\infty} d\tau
\cos
\left\{
(\epsilon_a+\epsilon_b-{\epsilon'_a}-{\epsilon'_b}){\tau \over \hbar} \right .
\nonumber\\&&
- \left .
\frac{{\bf E}\tau^2}{2 \hbar}
\left(
\frac{e_a{\bf k_a}}{m_a}+ \frac{e_b{\bf k_b}}{m_b}-
\frac{e_a {\bf k'_a}}{m_a}-\frac{e_b {\bf k'_b}}{m_b}
\right)
\right\}\nonumber\\
&&\times\left\{
f_{a'} f_{b'}(1-{f}_a)(1-{f}_b)-{f}_a {f}_b (1-f_{a'})(1-f_{b'})
\right\}
\nonumber\\
&&
\end{eqnarray}
describing the scattering of particles $a$ (electrons) with other
species $b$
with the distribution function $f_b=f_b(k_b-e_bE\tau,T-\tau)$. 
The potential turns out to be the static Debye one 
\be
V_s(q)={2\pi e_a e_b
\hbar\over(q+\hbar \kappa)}
\label{vs1}
\ee
with the static screening length $\kappa$ given by
\beq\label{screen}
\kappa=\sum\limits_c {2 \pi e_c^2 \partial_\mu n_c}
\label{kap}
\eeq
and the chemical potential $\mu$. 

We will now use this statically screened result in order to describe the scattering neutral impurities if we use the range of potential $r_0=1/\kappa$ and replace the charges by the scattering strength $g_{ab}=e_a e_b$. 

The calculation of the impurity scattering in quasi-two dimensions is
now analogously to the Brooks-Hearing result for three dimension 
and starts from the Born
collision integral (\ref{kinetic}) which takes for infinite heavy ions
[$m_b/m_a\rightarrow ]\infty$]
\be
&&I_{ab}(k_a) =
\!\frac{2 s_b}{\hbar^2}\!\int \!\! \frac{ d {\bf k'_a} d {\bf k_b} d {\bf k'_b}
}{(2\pi\hbar)^4}
\delta \left({\bf k_b}-{\bf k'_b}
\right )V_{s}^2({\bf k_a}\!-\! {\bf k'_a},t)
\nonumber\\
&&\times 
\int\limits_0^{\infty} d\tau
\cos
\left\{
(\epsilon_a-{\epsilon'_a}){\tau \over \hbar} 
- 
\frac{{\bf E}\tau^2}{2 \hbar}
\left(
\frac{e_a{\bf k_a}}{m_a}-
\frac{e_a {\bf k'_a}}{m_a}
\right)
\right\}
\nonumber\\
&&\times
f_{b}\left\{
f_{a'} -f_a \right\}.
 \nonumber\\
&&
\label{st1}
\ee
We assume parabolic bands $\epsilon=k^2/2m$. 

The relaxation function $\delta E/E$ would correspond to linearization
of the $cos$ function in (\ref{st1}) with respect to the field while
the relaxation time is obtained taking into account the linearization
with respect to the momentum $p_a$ of the displaced distributions (\ref{sh}). 
Cross terms like $p_a
E$ are already of second order response. The result can be written in the concise form (\ref{transi}).

\subsubsection{Relaxation time by neutral impurity scattering}

In the following we give an explicit calculation.
Employing the Yukawa or Debye potential (\ref{vs1}) one obtains
for the relaxation part
\be
&&I_{ab}^R(k_a)=-p_a {\partial f_0 \over \partial_{k_a}}
 {2 \pi m_a  \over \hbar}g_{ab}^2\int\limits_0^{\infty}
d\alpha {\cos{(\alpha-\phi)}-\cos{\phi} \over (2 k_a |\sin{\alpha\over
    2}|+\hbar \kappa)^2} 
\nonumber\\&&
\ee
with the angle between $k_a$ and the field direction $p_a$ denoted by $\phi$.

The current relaxation time is now obtained by
\be
n_a {\bf p_a} \tau_{ei}^{-1}=\int {d k_a \over (2 \pi \hbar)^2} {\bf
  k_a} I_{a b}(k_a)
\ee
from which one gets [$\kappa_p=\hbar /2 r_0 p$]
\be\label{taui}
\tau_i^{-1}&=&{m_a g_{ab}^2\over 2^{3/2} \hbar^3} {n_b s_b\over
  n_a}\int\limits_0^\infty dp \partial_p f_0 \left [{1\over \kappa_p^2-1}+{ \ln\left
  ({1+\sqrt{1-\kappa_p^2}\over \kappa_p}\right )\over (\kappa_p^2-1)^{3/2}}
\right ].
\nonumber\\&&
\ee
This current relaxation time in 
the low temperature Sommerfeld-expansion leads to
 [$\kappa_i=\hbar /2r_0 p_{fa}$]
\be
\tau_i^{-1}&=&{m_a g_{ab}^2\over 2^{3/2} \hbar^3} {n_b s_b\over n_a}\left (
{1\over \kappa_i^2-1}+{ \ln\left
  ({1+\sqrt{1-\kappa_i^2}\over \kappa_i}\right )\over (\kappa_i^2-1)^{3/2}}
\right .
\nonumber\\&&
\left . +{\pi^2\over 24}{T^2  \over \epsilon_f^2}\partial_{\kappa_i}^2  \left [
{1\over \kappa_i^2-1}+\kappa_i^4 { \ln\left
  ({1+\sqrt{1-\kappa_i^2}\over \kappa_i}\right )\over (\kappa_i^2-1)^{3/2}}
\right]\right ). 
\nonumber\\&&
\label{ti}
\ee
The second temperature correction is negative and diminishes the
positive first part. As long as $T<\epsilon_f$ the net relaxation time
is positive and continously falling to zero for $\kappa_i \to \infty$.

The high density expansion reads
\be
R_i=\tau_i^{-1} &&{m_a\over n_a e_a^2} {e_a^2\over h}={2^{3/2}s_b n_b\over
  s_a n_a}\left ({m_a g_{ab}
  r_0 \over \hbar^2}\right )^2 \kappa_i'^2 (1+\sqrt{\xi})^2
\nonumber\\&&\times \left
  \{
-1-\ln{\kappa_i\over 2}-(5+6 \ln{\kappa_i\over 2}) {\kappa_i^2\over 4}
\right .\nonumber\\&&\left .
-{\pi^2\over 24}{T^2  \over e_f^2}\left
  (2+(19+12\ln{\kappa_i\over 2})\kappa_i^2\right )
+o(\kappa_i^3)
\right \} 
\nonumber\\&&
\ee
and the low density expansion or short range expansion
\be
R_i&=&{2^{3/2}s_b n_b\over
  s_a n_a}\left ({m_a g_{ab}
  r_0 \over \hbar^2}\right )^2 \kappa_i'^2 (1+\sqrt{\xi})^2
\nonumber\\&&\times
\left
  \{
{1\over \kappa_i^2}-{\pi\over 2 \kappa_i^3}
-{\pi^2\over 24}{T^2  \over
    \epsilon_f^2}{\pi^3\over 16 \kappa_i^3}
+o(\kappa_i^{-4})
\right \} 
\nonumber\\&&
\label{til}
\ee

\subsubsection{Relaxation effect by neutral impurity scattering}

The relaxation function is now obtained if we linearize (\ref{st1})
with respect to the external field. We obtain
\be
&&\int {d k_a \over (2 \pi \hbar)^2} {\bf
  k_a} I_{a b}^{\delta E}(k_a)=-{\bf E} {n_i e\over m \hbar^3} \int {d k d q \over (2 \pi \hbar)^4}
f_0(\epsilon_k) V_s^2(q) q^2\nonumber\\&&
\times  \cos^2(q,E) \int\limits_0^\infty d \tau
\tau^2 \sin{\left ({q^2+2 {\bf k q}\over 2 m \hbar}\tau\right )}
\ee
from which one gets
\be
{\delta E\over E}&=&{n_i\over 4 \pi \hbar^5 m n_a} \int\limits_0^\infty
d q q^3 V_s^2(q) \int\limits_0^\infty d\tau \tau^2 I_s
\nonumber\\
&=&{n_i\over n_a} \kappa_a^2 \left ({m g_{ab} r_0\over \hbar^2}\right
)^2
\int\limits_0^\infty {d y\over (y+\kappa_a)^2} \partial_y^2
  \sqrt{y^2-1} \Theta(y-1)
\nonumber\\&&
\ee
where we have used (\ref{istt}) for the last line. 
Employing the same regularization due to the principal value,
(\ref{struc}), we end up with
\be
{\delta E\over E}&=&{n_i\over n_a}\left ({m g_{ab} r_0\over \hbar^2}\right
)^2 \left \{ {2 \kappa_a^2\!+\!\kappa_a^4\over (\kappa_a^2\!-\!1)^2}\!+\!{3
    \kappa_a^3 \ln{(\kappa_a\!-\!\sqrt{\kappa_a^2\!-\!1})}\over
    (\kappa_a^2\!-\!1)^{5/2}}  \right\}.
\nonumber\\&&
\label{imprel}
\ee
The $T^2$ dependent term could be given analogously.
The needed low density or short range expansion reads now
\be
{\delta E_i\over E}&=&{n_i\over n_a}\left ({m g_{ab} r_0\over \hbar^2}\right
)^2 \left \{ 1+{4-3\ln{2\kappa_a}\over \kappa_a^2} \right\}+o(\kappa^{-4}).
\label{imprel1}
\ee

\section{Polarization function in 2D}\label{pol2D}
Here we discuss the properties of low temperature polarization function
\be
\Pi(\omega,q)&=&\int {d p \over (2 \pi \hbar)^2} {f_0({(p+{q\over
    2})^2\over 2 m})-f({(p-{q\over 2})^2 \over 2 m})
\over {p\cdot q\over m}-\omega-i 0}.
\label{pi}
\ee
The imaginary part is easily rewritten as
\be
&&{\rm Im} \Pi(\omega,q)=\pi \int\limits_0^ \infty {d p p \over (2 \pi
  \hbar)^2}f_0({p^2\over 2
  m})\int\limits_0^ {2 \pi} d \phi 
\nonumber\\&&\times 
\left \{ \delta ({p.q\over
    m}-{q^2\over 2 m}-\omega)-\delta({p.q\over m}+{q^2\over 2
    m}-\omega) \right \}
\nonumber\\&=&
{m^{3/2} \over 2^{3/2} \pi \hbar^2 q}
\int\limits_0^\infty {d \epsilon \over \sqrt{\epsilon}} \left (f_0(\epsilon+{(q/2+m \omega
    \hbar/q)^2\over 2 m})\right .
\nonumber\\&&
\left .-f_0(\epsilon+{(q/2-m \omega
    \hbar/q)^2\over 2 m}) \right ).
\nonumber\\&&
\label{imp}
\ee
The energy shifts in the distribution function we absorb into an
effective chemical potential which should be positive in order to
obtain nonzero contribution at Sommerfeld expansion
\be
\mu_{\rm eff}=\epsilon_f -{\left (\pm {q\over 2}+{m \omega \hbar \over
      q}\right )^2
  \over 2 m}\ge0
\ee
The low temperature Sommerfeld expansion reads than
$f_0(\epsilon)=n((\epsilon-\mu_{\rm eff})/T)$
\be
&&\int\limits_0^\infty {d \epsilon\over \sqrt{\epsilon}}
n({\epsilon-\mu_{\rm eff}\over T})
=2 \int\limits_{\mu_{\rm eff}/T}^\infty dx  n(x) \partial_x \sqrt{T
  x+\mu_{\rm eff}}
\nonumber\\&&\qquad\qquad=
2 \int\limits_{\mu_{\rm eff}/T}^\infty d x n(x)(1-n(x)) \sqrt{T
  x+\mu_{\rm eff}}
\nonumber\\&&\qquad\qquad=2 \sqrt{\mu_{\rm eff}} \left (1-{\pi^2 T^2\over 24 \mu_{\rm eff}}
  \right ).
\nonumber\\&&\qquad\qquad=2\left (1-{\pi^2 T^2\over 12}{\partial\over
  \partial \epsilon_f}\right ) \sqrt{\mu_{\rm eff}}.
\ee
Therefore it is enough to know the zero temperature result since the
$T^2$ correction are given simply by derivatives with respect to $\epsilon_f$.

\begin{figure}
\noindent\psfig{file=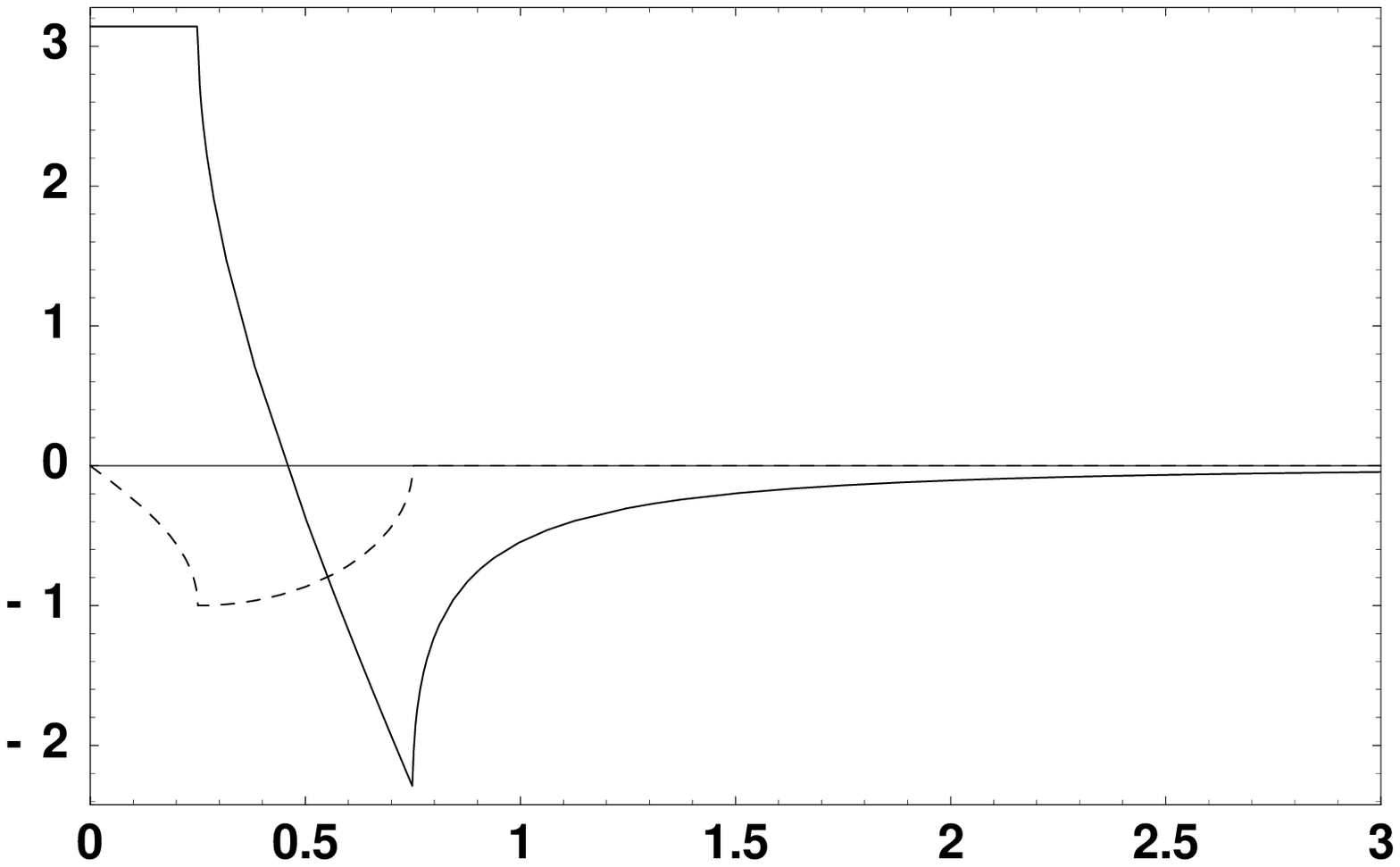,width=8cm}
\psfig{file=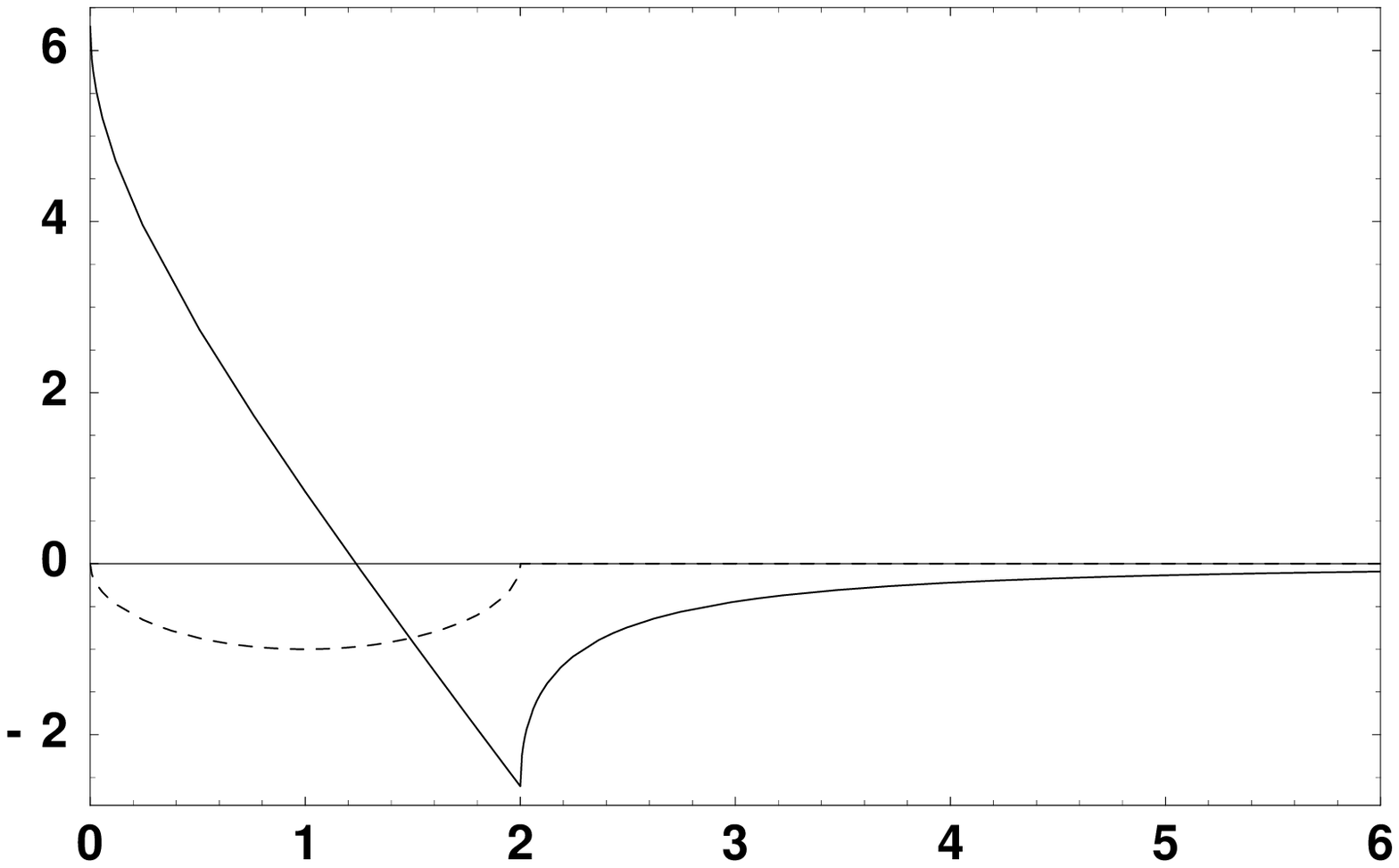,width=8cm}
\psfig{file=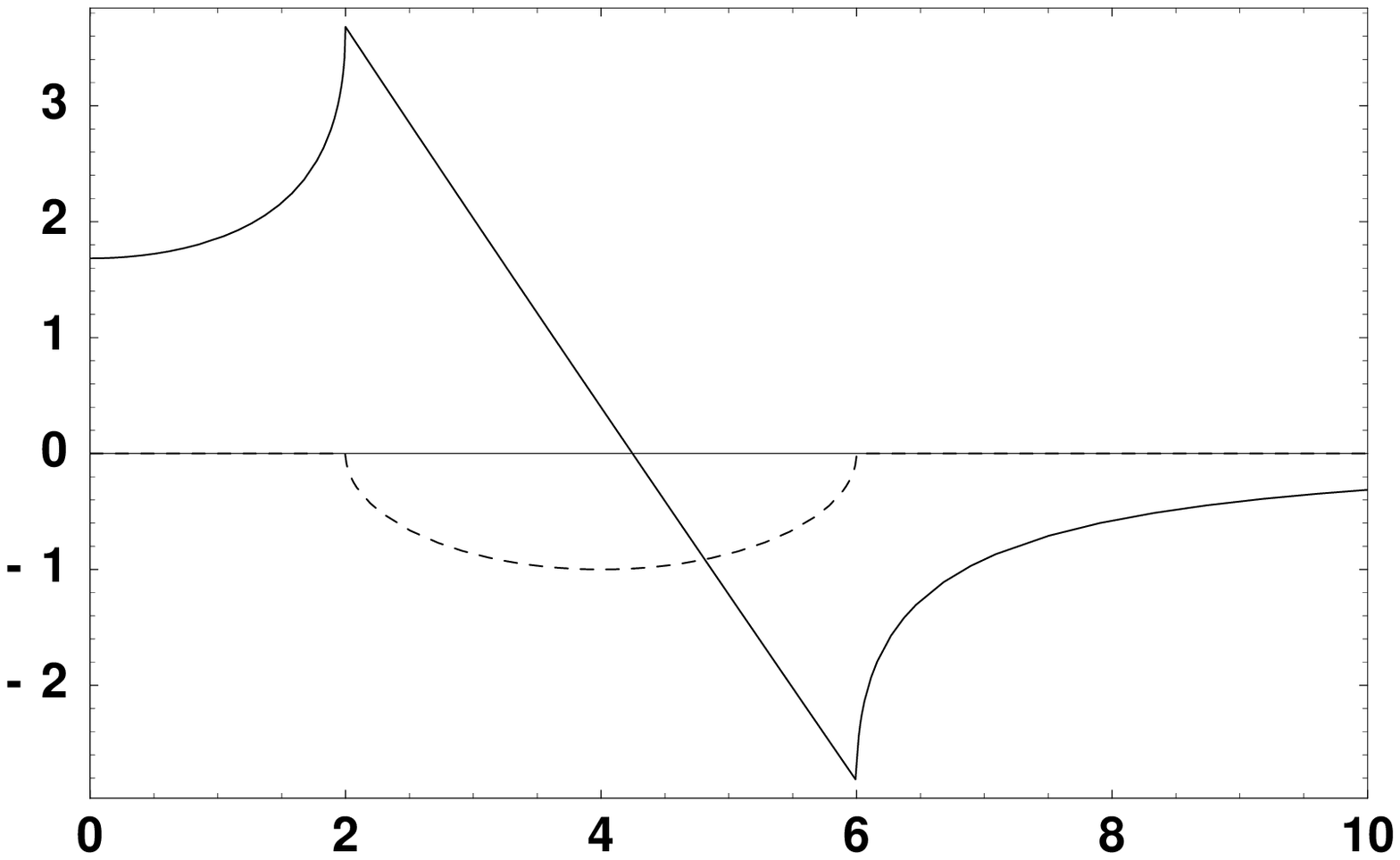,width=8cm}
\caption{The real (solid) and imaginary (dashed) part of the
  polarization $4 \pi \hbar^2/m\times \Pi$ versus $x_0$ for $x=0.5$
  (above), $x=1$ (middle) and $x=2$ (below) according to
  (\protect\ref{cco}).\label{reim2d}}
\end{figure}

Introducing dimensionless coordinates as in \cite{HSW71} 
\be
x={q\over 2 p_f} \qquad x_0 ={\hbar \omega \over 4 \epsilon_f}
\label{cco}
\ee
we get finally for the imaginary part of the polarization function
\be
&&{\rm Im}\Pi(\omega,q)={m\over 4 \pi \hbar^2 x} 
\nonumber\\&&\times\left \{ \Theta(x-|x_0+x^2|) \sqrt{1-\left (x+{x_0\over x} \right )^2}
\right .\nonumber\\&&\left .- \Theta(x-|x_0-x^2|) \sqrt{1-\left (x-{x_0\over x} \right )^2}\right \}
\label{impf}
\ee
which is of course the result given in \cite{HSW71}.
The corresponding real part is given by the Hilbert transform
according to (\ref{pi})
\be
{\rm Re}\Pi(\omega,q)=-2 \int {d\omega'\over 2 \pi} {{\rm
    Im}\Pi(\omega',q)\over \omega-\omega'}.
\ee
Using the integral
\be
f(a)=\int\limits_{-1}^1 {\sqrt{1-z^2}\over a-z}=\pi \left \{\begin{array}{ll} a&1\ge |a|\cr a-{\rm
    sgn}(a) \sqrt{a^2-1}\end{array} \right .
\nonumber\\&&
\ee
we obtain
\be
{\rm Re} \Pi=-{m\over 4 \pi^2 \hbar^2 x} \left \{f({x_0\over
    x}+x)-f({x_0\over x}-x)\right \}.
\label{repf}
\ee
The real and imaginary part is plotted in figure~\ref{reim2d}.

\section{Integrals over dielectric functions}\label{a1}

In order to perform the frequency integration in (\ref{kin1}) 
we use a very useful relation, which has been given in \cite{K75}
\beq
&&\int \!\!{d \omega \over 2 \pi} {H(\omega) \over \omega} {\rm Im}
{\cal E}^{-1}(q,\omega)\!=\!\frac{H(0)}{2} {\rm Re} \left ( \!1\!-\!{1
    \over {\cal E}(q,0)}
\!\right ) \label{a13}.
\nonumber\\
&&
\eeq

For the integration of (\ref{kin1}) we set $H(\omega)=\cos{(\omega \tau+A)}\cos{(\omega\tau'+B)}
\omega/{\rm Im} {\cal E}$ with $A$ and $B$ are the remaining content of the cosine
functions of (\ref{eqe}) and (\ref{fluc1}). 

Lets first prove the relation (\ref{a13}). We consider the following
integral including the dielectric function
\beq\label{a10}
I&=&\int {d \omega \over 2 \pi} {H(\omega) \over \omega} {\rm Im }
{\cal E}^{-1}(\omega)\nonumber\\
&=&\int \! {d \omega \over 4 \pi i} \!\left (\!{1\over \omega +i \eta} \!+\!
{1\over \omega -i \eta}\! \right ) H(\omega) (f^-\!-\!f^+)
\eeq
where $f^+=1-1/{\cal E}$ and $f^-=(f^+)^*$. In the following we will assume that the function $H(\omega)$ is analytical.
Since $f^{\pm}(\omega)$ has no poles in the lower/upper half plane and
vanishes with $\sim \omega^{-2}$ for large $\omega$ we have the identity
\beq
\int {d \omega \over 2 \pi i} H(\omega) {f^{\pm}(\omega) \over (\omega \pm i \eta)} =\mp f^{\pm}(0) H(0)\label{a11}
\eeq
and all other combinations of signs vanish. 
With the help of the relation (\ref{a11}) we compute easily for (\ref{a10})
\beq
I=\frac 1 2 H(0) {\rm Re} \left ( 1-{1 \over {\cal E}(0)}\right )
\label{fina}
\eeq
which proves relation (\ref{a13}).

\subsection{Regularization of integration}
We are now going to give explicit forms including the dielectric function ${\cal E}=1-V(q) \Pi(q,\omega)$ where the
polarization function $\Pi$ was discussed in the foregoing chapter.
The forms appearing throughout the paper are 
\beq\label{a12}
I&=&\int {d \omega \over 2 \pi} {h(\omega) \over
|{\cal E}(\omega)|^2}\nonumber\\
&=&-\int {d \omega \over 2 \pi} {\omega h(\omega)/{\rm Im}{\cal E}(\omega) \over \omega} {\rm Im }
{\cal E}^{-1}(\omega)
\eeq
such that we can apply (\ref{fina}) with $H(\omega)=\omega
h(\omega)/{\rm Im}{\cal E}(\omega)$. In the case where
${\rm Im}{\cal E}= 0$ appears an ambiguity which we  have to remove.
We add to the particle-hole
fluctuation which forms the polarization an infinitesimal small 
classical process
\be
\delta {\rm Im}\Pi=\eta \omega {\rm e}^{-c \omega^2}\approx \eta (\omega+o(\omega^3)).
\ee
This will make the imaginary part of the polarization nonzero
everywhere and will not introduce additional poles in the upper half
plane. Therefore we can apply the integration (\ref{fina}). 
The corresponding real part according to (\ref{fina}) will be $\propto
\eta$ and drop out in the final form (\ref{fina}). 

\subsection{Specific forms}
Now we write down the required forms for (\ref{fina}).
One gets the static result from (\ref{impf}) and (\ref{repf})  at low temperatures \cite{HSW71}
\be
\Pi(0,q)=-{\partial \over \partial \mu} n \left
  \{\begin{array}{ll}1&q<2p_f\cr 1-\sqrt{1-\left ({2 p_f\over q}\right
      )^2} & q>2 p_f
\end{array}\right .
\nonumber\\&&
\label{p0}
\ee
with the chemical potential $\mu$ and from (\ref{imp})
\be
&&{\rm Im} \Pi(\omega,q)=\nonumber\\&&-{m^2 \omega \hbar \Theta(2 p_f-q)\over \pi \hbar^2 q \sqrt{4 p_f^2-q^2}} \left
    (1+{8\pi^2 T^2 m^2 \over (4 p_f^2-q^2)^2} \right )\nonumber\\&&+o(\omega^2,T^4)
\nonumber\\&&
\label{imp2}
\ee
The region where ${\rm Im}\Pi\ne 0$ correspond exactly to 
the upper case of (\ref{p0}). 
Using this expansion we obtain
\be
&&H(0)=\lim\limits_{\omega \to 0} {\hbar \omega \over {\rm Im}{\cal E^R}}=
-{\hbar q^2 \over \sum\limits_b { m_b^2 e_b^2\over \sqrt{4
    p_{fb}^2-q^2}} (1+{8 \pi^2 m_b^2 T_b^2\over (4 p_{fb}^2-q^2)^2})}
\nonumber\\&&
\ee
where we have to keep in mind that the above procedure in calculating
the frequency integral works only for finite ${\rm Im}\Pi$ or finite
particle-hole fluctuations. According to (\ref{imp}) this restricts
the 
later $q$-integration
to values smaller than $2p_{fb}$ respectively. 

We get finally for (\ref{a13})
\be
W(q)&=&V(q)^2 \frac 1 2 \lim\limits_{\omega \to 0} {\hbar \omega \over
  {\rm Im}{\cal E^R}} {\rm Re} (1-{1\over {\cal E}^R(0,q)})
\nonumber\\
&=&-{2 \pi e_a^2 e_b^2 \hbar^4 \kappa(q)\over q+\hbar \kappa(q)}{1 \over \sum\limits_b { m_b^2 e_b^2\over \sqrt{4
    p_{fb}^2-q^2}} (1+{8 \pi^2 m_b^2 T_b^2\over (4 p_{fb}^2-q^2)^2})}
\nonumber\\&&
\label{W}
\ee 
where the screening length is from (\ref{p0})
\be
\kappa &=&\sum\limits_b 2 \pi
e_b^2\partial_\mu n_b \left
  \{\begin{array}{ll}1&q<2p_{fb}\cr 1-\sqrt{1-\left ({2 p_{fb}\over q}\right
      )^2} & q>2 p_{fb}
\end{array}\right .\nonumber\\
\label{kq}
\ee
and since $n={s\over 4 \pi \hbar^2} p_f^2 +o({\rm e}^{-\epsilon_f/T})$ one has
$\partial_\mu n= {m_b s_b \over 2 \pi \hbar^2}$.

\section{Low temperature expansion of Integrals}\label{integral}

The integrals occurring in (\ref{relaxf}), (\ref{rf}) and (\ref{isa}) will now be
calculated. Using $\int\limits_0^ \infty d \tau \cos{x\tau}=\pi
\delta(x)$ we can write for (\ref{fluc}) and (\ref{po})
\be
&&-\frac 1 2 \Pi^<(q,\omega=0)=\int {d p\over (2 \pi\hbar)^2} f_b(p)(1-f_b(p+q)) 
\nonumber\\ &&\qquad \qquad\times \int\limits_0^\infty d \tau' \cos{\left (
    \epsilon_{p+q}-\epsilon_p \right )
{\tau'\over \hbar}}
\nonumber\\&&
=\int {d p\over (2
  \pi\hbar)^2} f_b(p)(1-f_b(p)) \pi \hbar \delta\left (
    \epsilon_{p+q}-\epsilon_p \right )
\nonumber\\
&&={m\over 4 \pi \hbar q}\int\limits_0^\infty d p
f_b(p)(1-f_b(p))\int\limits_{-1}^1 {d x \over \sqrt{1-x^2}}
\nonumber\\&&\qquad \qquad\times 
\left (\delta
  (x+{q\over 2 p})+\delta
  (x-{q\over 2 p})\right ) 
\nonumber\\
&&={m_b^2 T_b\over \pi \hbar q}\int\limits_{{q^2\over 8 m_b T_b}-{\epsilon_{fb}\over T_b}}^\infty d x
{n(x)(1-n(x))\over \sqrt{8 m_b (T_b x+\epsilon_{fb})-q^2}} 
\label{potr}
\ee
where we used $x=p^2/2m_bT_b-\epsilon_{fb}/T_b$ and $n(x)=1/({\rm
    e}^x+1)$ and $\epsilon_{fb}$ is the Fermi energy. 
The last integral is only non-zero for negative lower integration
limits implying $2 p_{fb}>q$. Expanding the square root in terms of the temperature
  $T_b$ we obtain finally
\be
&&-\frac 1 2 \Pi^<(q,\omega=0)=-\frac 1 2 \Pi^>(q,\omega=0)=
\nonumber\\&&{m_b^2 T_b \Theta(2 p_{fb}-q) \over \pi \hbar q
  \sqrt{4 p_{fb}^2-q^2}} \left ( 1+{8 \pi^2 m_b^2 T_b^2 \over (4
      p_{fb}^2-q^2)^2}\right ) \nonumber\\&&
\label{ic}
\ee

The integrals (\ref{relaxf}), (\ref{rf})  can be tremendously
simplified observing that (\ref{isa}) can be written
\be
&&I_s(a,\tau)=\int dk f_k(1-f_{k-q}) \sin{\left (
    \epsilon_{k-q}-\epsilon_k \right ){\tau\over \hbar}}
\nonumber\\
&=&\int dk (f_{k-q}-f_k) g_{\epsilon_{k-q}-\epsilon_k} \sin{\left (
    \epsilon_{k-q}-\epsilon_k \right ){\tau\over \hbar}}
\nonumber\\
&=&\int dk f_k \left (g_{\epsilon_{k+q}-\epsilon_k} \sin{\left (
    \epsilon_{k}-\epsilon_{k+q} \right ){\tau\over \hbar}}\right .
\nonumber\\&&\qquad \qquad \left . -
g_{\epsilon_{k}-\epsilon_{k-q}} \sin{\left (
    \epsilon_{k-q}-\epsilon_{k} \right ){\tau\over \hbar}}\right )
\nonumber\\
&=&\int dk f_k \sin{\left (
    \epsilon_{k}-\epsilon_{k-q} \right ){\tau\over \hbar}} \left (g_{\epsilon_{k-q}-\epsilon_k} +
g_{\epsilon_{k}-\epsilon_{k-q}} \right )
\nonumber\\
&=&\int dk f_k \sin{\left (
    \epsilon_{k-q}-\epsilon_{k} \right ){\tau\over \hbar}} 
\ee
where $g_x=1/({\rm e}^{x/T}-1)$ and we have used  $k\to -k$
transformation in coming from the third to the fourth equality.

The time integral (\ref{rf}) can be now represented as a derivative
of a $\delta$ function with respect to $k$. A partial integration
leads than to
\be
&&\int\limits_0^\infty d \tau \tau I_s(a,\tau)=\int {dk\over (2 \pi \hbar)^2} f_k
\int\limits_0^\infty d \tau \tau \sin{\left (
    \epsilon_{k-q}-\epsilon_{k} \right ){\tau\over \hbar}}
\nonumber\\&&
=-{m_a^2 \over 4 \pi q^2}\int\limits_{q/2}^\infty {d k\over k}
\partial_k (k f_a(k))\int\limits_{-1}^1 {d x \over x \sqrt{1-x^2}}
\nonumber\\&&\qquad\qquad\times 
\left (\delta
  (x-{q\over 2 p})-\delta
  (x+{q\over 2 p})\right ) 
\nonumber\\&&
=-{m_a^2 \over \pi q^3}\int\limits_{q/2}^\infty {d k}
\partial_k (k f_a(k)){k \over \sqrt{k^2-{q^2\over 4}}}.
\nonumber\\&&
={m_a^{3/2} \over 4 \sqrt{2}\pi q }\int\limits_{-a/T_a}^\infty d
  x n(x) (1-n(x))
{\Theta(2 p_{fa}-q)\over \sqrt{T_ax+a}}
\ee
with $a=\epsilon_{fa}-q^2/8m_a$. Expanding
the argument in terms of $T_a$ one arrives analogously to (\ref{ic})
at the result
\be
&&\int\limits_0^\infty d \tau \tau I_s(a,\tau)=
{m_a^2  \over 4 \pi  q
  \sqrt{4 p_{fa}^2-q^2}} \left ( 1+{8 \pi^2 m_a^2 T_a^2 \over (4
      p_{fa}^2-q^2)^2}\right ) \nonumber\\&&\times\Theta(2 p_{fa}-q).
\label{ist}
\ee

The integral (\ref{relaxf}) requires some more care. We use
the principal value identity $\int\limits_0^\infty d\tau \sin{\tau
  x}={\cal P}/x$ and write
\be
&&\int\limits_0^\infty d \tau \tau^2 I_s(a,\tau)=
\nonumber\\&&
-{m_a^3 \hbar \over \pi^2 q^3}{\cal P}\int\limits_{0}^\infty {d
  k}f_a(k)\partial_q^2 \int\limits_0^{2\pi} {d\phi\over {q\over 2
    k}-\cos\phi}.
\ee
Using 
\be
\int\limits_0^{2\pi} {d\phi\over a-\cos\phi}={2 \pi\over
    \sqrt{a^2-1}}\Theta(a-1)
\ee
we can write after one partial integration
\be
&&\int\limits_0^\infty d \tau \tau^2 I_s(a,\tau)=
\nonumber\\&&
{m_a^3 \hbar {\cal P}\over \pi q^3}\partial_q^2 \int\limits_{-\epsilon_{fa}/T}^{q^2/8m_aT_a-\epsilon_{fa}/T_a} {d
  x}n(x) (1-n(x))\nonumber\\&&\times\sqrt{q^2-8 m_a T_a x -4 p_{fa}^2}.
\ee
Expanding again the argument in terms of $T_a$ we arrive at
\be
&&\int\limits_0^\infty d \tau \tau^2 I_s(a,\tau)={m_a^3 \hbar {\cal P} \over
  \pi q^3}\nonumber\\&&\times{\partial^2\over \partial q^2}\left (
  \sqrt{q^2-4 p_{fa}^2} \left ( 1-{8 \pi^2 m_a^2 T_a^2 \over 3 (
      q^2-p_{fa}^2)^2}\right ) \Theta(q-2 p_{fa})\right ).
\label{istt}
\nonumber\\&&
\ee


\end{document}